\documentclass[12pt]{article}

\begin{document}
\begin{center}
 {\bf{\Large Einstein's Field Equations for the Interior of a Uniformly
 Rotating Stationary Axisymmetric Perfect Fluid}}
\vspace{1cm}

E. Kyriakopoulos\\
Department of Physics\\
National Technical University\\
157 80 Zografou, Athens, GREECE
\end{center}

\begin {abstract}
We reduce Einstein's  field equations for the interior of a
uniformly rotating, axisymmetric perfect fluid to a system of six
second order partial differential equations for the pressure $p$
the energy density $\mu$ and four dependent variables.Four of
these equations do not depend on $ p $ and $\mu$ and the other two
determine $ p $ and $\mu$.

PACS number(s): 04.20.Jb, 04.20.-q
\end {abstract}

\section{Introduction}
   The number of solutions of Einstein's field equations for the
interior of a uniformly rotating stationary axisymmetric  perfect fluid is very
limited contrary to what happens in the stationary axisymmetric
vacuum case \cite{kr:1}.This is mainly due to the fact that
the equations we have to solve in the first case are
complicated \cite{tr:2}, while in the second case they have been
reduced to the equivalent and much simpler Ernst's
equation \cite{er:3}.Reduction of the "interior" problem is known
for Einstein-Mawxell field \cite{er:4} and dust \cite{wi:5}.Also
Einstein's field equations for the interior of a uniformly
rotating stationary axisymmetric perfect fluid have been reduced
to a system of two second order partial differential equations for
two unknown functions, which however are very
complicated \cite{bo:6}.

In this work the six Einstein's field equations for the interior
of a uniformly rotating stationary axisymmetric perfect fluid,
which depend on the pressure $ p $ the energy density $\mu$ of the
fluid and on four dependent variables, will be divided into a set
of four equations which depend on the four  dependent variables
but not on $ p $ and $\mu$ and a set of two equations which give
m$ p $ and $\mu$ as functions of the four dependent variables.
Therefore solving the first system of equations we can determine $
p $ and $\mu$ algebraically using the second set. Introducing a
potential and redefining variables we write the first system in a
relatively simple form, which is very convenient for finding the
B\"{a}cklund transformations the equations of the system may
posses. Also we eliminate one of the dependent variables of the
system increasing however the complexity of the problem.

\section{The Field Equations}
 The line element of stationary axisymmetric fields admitting
 2-spaces orthogonal to the Killing vectors
$ \vec{\xi} =\partial_t$ and  $\vec{\eta} =\partial_\phi$ can be
 written in the form \cite{eh:7}, \cite{kr:1}
\begin{equation}
ds^{2}=e^{-2U}\{e^{2K}(d\rho^{2}+dz^{2})+
F^{2}d\phi^{2}\}-e^{2U}(dt+Ad\phi)^{2}           \label{21}
\end{equation}
where $U=U(\rho,z)$, $K=K(\rho,z)$, $F=F(\rho,z)$ and
$A=A(\rho,z)$. The decomposition of the metric into orthogonal
2-spaces whose points are labeled by $\rho$ and z is possible for
perfect fluid solutions provided that the 4-velocity of the
fluid satisfies the circularity condition
$u_{[a}\xi_{b}\eta_{c]}=0$ \cite{kr:1}.

For a perfect fluid source we have the energy-momentum tensor
\begin{equation}
 T_{ab}=(\mu+p)u_{a}u_{b}+pg_{ab}  \label{22}
\end{equation}
where $u_{a}$ are the components of the 4-velocity and $\mu$ and $
p $ are the mass density and the pressure of the fluid
respectively. We shall introduce the notation
\begin{equation}
\partial_{\rho}=\frac{\partial}{\partial_{\rho}} \mbox{,  }
\partial_{z}=\frac{\partial}{\partial_{z}} \mbox{,  }
\partial=\frac{1}{2}(\partial_\rho-i\partial_{z}) \mbox{,  }
\overline{\partial}=\frac{1}{2}(\partial_{\rho}+i\partial_{z}).
\label{23}
\end{equation}
Then Einstein's field equations for a uniformly rotating
axisymmetric perfect fluid can be written in the form \cite{tr:2}
\begin{eqnarray}
&&2\partial\overline{\partial}F=pFe^{2K-2U} \label{24}\\
&&2\partial\overline{\partial}U+\frac{1}{F}(\partial
U\overline{\partial}F+ \overline{\partial}U\partial F)+
\frac{1}{F^{2}}e^{4U}\partial A\overline{\partial}A
=\frac{1}{4}(\mu+3p)e^{2K-2U} \label{25}\\
&&2\partial\overline{\partial}A-\frac{1}{F}(\partial
A\overline{\partial}F+ \overline{\partial}A\partial F)+4(\partial
A\overline{\partial }U+ \overline{\partial}A\partial U)=0
\label{26}\\ && 2\overline{\partial}F\overline{\partial}K-
\overline{\partial}\,\overline{\partial}F-
2F\overline{\partial}U\overline{\partial}U+
\frac{1}{2F}e^{4U}\overline{\partial}A\overline{\partial}A=0
\label{27}\\ &&\partial\overline{\partial}K+\partial
U\overline{\partial}U + \frac{1}{4F^{2}}e^{4U}\partial
A\overline{\partial}A= \frac{1}{4}p e^{2K-2U} \label{28}
\end{eqnarray}
Also from the conservation relation $T^{ab}_{;b}=0$ we get
equations
\begin{equation}
\partial p+(\mu+p)\partial U=0  \label{29}
\end{equation}
which are obtained from Eq. (\ref{24})  -  (\ref{28})  and will be omitted in
the following.

Eqs  (\ref{24}), (\ref{25})  and  (\ref{28})  are equivalent to the system
\begin{eqnarray}
&&p=\frac{2}{F}\partial\overline{\partial}Fe^{2U-2K} \label{210}\\
&&\mu=e^{2U-2K}\{8\partial\overline{\partial}U+\frac{4}{F}(\partial
U\overline{\partial}F+ \overline{\partial}U\partial
F)+\frac{4}{F^{2}}e^{4U}\partial A\overline{\partial}A\nonumber\\
&&-12\partial\overline{\partial}K- 12\partial
U\overline{\partial}U- \frac{3}{F^{2}}e^{4U}\partial
A\overline{\partial}A\}\label{211}\\
&&4\partial\overline{\partial}K+4\partial U\overline{\partial}U+
\frac{1}{F^{2}}e^{4U}\partial A\overline{\partial}A-
\frac{2}{F}\partial\overline{\partial}F=0 \label{212}
\end{eqnarray}
We shall replace Eqs  (\ref{24}),  (\ref{25})  and  (\ref{28}) by Eqs
 (\ref{210})  -  (\ref{212}).

If we put
\begin{eqnarray}
A=-\omega \mbox{, } e^{2U}=T    \label{213}
\end{eqnarray}
Eq  (\ref{26})  takes the form
\begin{equation}
\partial(\frac{T^{2}}{F}\overline{\partial}\omega)+
\overline{\partial}(\frac{T^{2}}{F}\partial\omega)=0 \mbox{ or   }
\partial_{\rho}(\frac{T^{2}}{F}\omega_{\rho})+\partial_{z}(\frac{T^{2}}{F}\omega_{z})=0
\label{214}
\end{equation}
where a letter as an index in a function means differentiation
with respect to the corresponding variable e.g.
$\omega_{\rho}=\frac{\partial \omega}{\partial \rho}$. The
solution of the above equation is
\begin{equation}
\frac{T^{2}}{F}\partial \omega=i\partial \phi  \label{215}
\end{equation}
or
\begin{eqnarray}
\omega_{\rho}=\frac{F}{T^{2}}\phi_z \mbox{ and }
\omega_z=-\frac{F}{T^{2}}\phi_\rho  \label{216}
\end{eqnarray}
where $\phi$ is an arbitrary function.
Also we have
\begin{eqnarray}
T\{\phi_{\rho\rho}+\phi_{zz}+\frac{1}{F}(F_{\rho}\phi_{\rho}+F_{z}\phi_{z})\}
- 2(T_{\rho}\phi_{\rho}+T_{z}\phi_{z})\nonumber\\
=\frac{T^{3}}{F}\{\partial_\rho(\frac{F}{T^{2}}\phi_{\rho})+
\partial_{z}(\frac{F}{T^{2}}\phi_{z})\}  \label{217}
\end{eqnarray}
But if Eqs  (\ref{216})  are satisfied the above relation vanish.
Therefore we can substitute Eq  (\ref{214})  by the relation
\begin{equation} \label{218}
T\{\phi_{\rho\rho}+\phi_{zz}+\frac{1}{F}(F_{\rho}\phi_{\rho}+F_{z}\phi_{z})\}-
2(T_{\rho}\phi_{\rho}+T_{z}\phi_{z})=0
\end{equation}
and Eqs  (\ref{216})  which define $\omega$.Then using Eqs  (\ref{216})  to
eliminate $\omega$ and introducing the Ernst's potential
\begin{equation}
E=T+i\phi \label{219}
\end{equation}
Eqs  (\ref{27}),  (\ref{210})  -  (\ref{212})  and   (\ref{218})  become
\begin{eqnarray}
&&Im\{\frac{1}{2}(E+\overline{E})(\nabla^{2}E+
\frac{1}{F}\vec{\nabla}F\cdot\vec{\nabla}E)  -
\vec{\nabla}E\cdot\vec{\nabla}E\}=Im{\Lambda}=0 \label{220}\\
&&2\overline{\partial}F\overline{\partial}K-
\overline{\partial}\,\overline{\partial}F -
\frac{2F}{(E+\overline{E})^{2}}
\overline{\partial}E\overline{\partial}\,\overline{E}=0
\label{221}\\
&&2\partial\overline{\partial}K-\frac{1}{F}\partial\overline{\partial}F+
\frac{1}{(E+\overline{E})^{2}}(\partial
E\overline{\partial}\,\overline{E}+
\overline{\partial}E\partial\overline{E})=0  \label{222}\\
&&p=\frac{1}{F}(E+\overline{E})e^{-2K}\partial\overline{\partial}F
\label{223}\\
 && \mu=e^{-2K}(\frac{2}{E+\overline{E}}Re\Lambda-N)=
-3p+\frac{2}{E+\overline{E}}e^{-2K}Re\Lambda  \label{224}
\end{eqnarray}
where
\begin{eqnarray}
&&\nabla^{2}=\partial^{2}_{\rho\rho}+\partial^{2}_{zz} \mbox{,  }
\vec{\nabla}=\hat{\rho}\partial_{\rho}+\hat{z}\partial_{z}
\label{225}\\ &&\Lambda=\frac{1}{2}(E+\overline E)(\nabla^{2}E+
\frac{1}{F}\vec{\nabla}F\cdot\vec{\nabla}E)-
\vec{\nabla}E\cdot\vec{\nabla}E  \label{226}\\
&&N=6(E+\overline{E})\partial\overline{\partial}K+\frac{3}{E+
\overline{E}}(\partial E\overline{\partial}\,\overline{E}+
\overline{\partial}E\partial\overline{E})=3pe^{2K}  \label{227}\
\end{eqnarray}
The expressions $Im\Lambda$ and $Re\Lambda$ are the imaginary part
and the real part of $\Lambda$ respectively, and $ \hat{\rho} $
and $ \hat{z} $ are the unit vectors in the direction of the
$\rho$ and z axis respectively. Also in deriving Eq  (\ref{227})
we have used Eqs  (\ref{222})  and  (\ref{223}). Thus we have
reduced the solution of the problem to the solution of Eqs
(\ref{220}) -  (\ref{222}),  which form a system of four equations
for the four unknowns $F$, $K$, and $E=T+i\phi $. Having solved
this system we can use Eqs  (\ref{223})  and  (\ref{224})  to
calculate $ p $ and $\mu$.

Our system of equations  (\ref{220}) -  (\ref{224})  is simplified
considerably in the vacuum case. In this case we have $p=\mu=0$
and we can introduced  Weyl's canonical coordinates with $ F=\rho
$ . Then Eq  (\ref{223})  is satisfied and Eq   (\ref{224})
implies the relation $ Re\Lambda$=0. This combined with Eq
(\ref{220}) gives
\begin{equation}
\Lambda=\frac{1}{2}(E+\overline{E})(\nabla^{2}E+\frac{1}{\rho}E_{\rho})-
\vec{\nabla}E\cdot\vec{\nabla}E=0  \label{228}
\end{equation}
which is Ernst's equation. Also Eqs  (\ref{221})  imply the relations
\begin{equation}
K_{\rho}=\frac{\rho}{4T^{2}}(T^{2}_{\rho}+\phi^{2}_{\rho}-T^{2}_{z}-\phi^{2}_{z})
\mbox{,  } K_{z}=\frac{\rho}{2T^{2}}(T_{\rho}T_{z}+\phi_{\rho}\phi_{z})
\label{229}
\end{equation}
which are consistent (i.e. $K_{\rho z}=K_{z \rho}$ )  if
$\Lambda=0$. Finally Eq  (\ref{222})  is a consequence of Eqs
(\ref{228}) and  (\ref{229}).  Therefore in the vacuum case we
have to solve Eqs  (\ref{228})  and  (\ref{229})  only.

If we put
\begin{equation}
lnF= G  \label{230}
\end{equation}
we get
\begin{equation}
\frac{1}{F}\partial_a{\partial_b}F=\partial_a{\partial_b}G+{\partial_a}G{\partial_b}G
\label{231}
\end{equation}
Then Eqs  (\ref{221}) and  (\ref{222})  become
\begin{eqnarray}
&&\overline{\partial}\,{\overline{\partial}}G+
{\overline{\partial}}G{\overline{\partial}}G -
2{\overline{\partial}}G{\overline{\partial}}K+
\frac{2}{(E+\overline{E})^{2}}{\overline{\partial}}E{\overline{\partial}}E=0
\label{232} \\
&&2\partial{\overline{\partial}}K-\partial{\overline{\partial}}G-
\partial G{\overline{\partial}}G+
\frac{1}{(E+\overline{E})^{2}}(\partial
E{\overline{\partial}}\,{\overline E} +
{\overline{\partial}}E{\partial}{\overline{E}})=0 \label{233}
\end{eqnarray}
To simplify the notation we shall introduce the variables $\xi$
and $\eta$ by the relations
\begin{eqnarray}
\xi=\rho+iz \mbox{ and } \eta=\rho-iz  \label{234}
\end{eqnarray}
and we shall put
\begin{equation}
2K-G=M  \label{235}
\end{equation}
Then Eqs  (\ref{220}), ( \ref{232})  and  (\ref{233})  become
\begin{eqnarray}
&&G_{\eta\eta}-M_{\eta}G_{\eta}+\frac{1}{2 T^{2}}(T^{2}_{\eta}+
\phi^{2}_{\eta})=0 \label{236}\\
&&M_{\eta\xi}-G_{\eta}G_{\xi}+\frac{1}{2 T^{2}}(T_{\eta}T_{\xi}+
\phi_{\eta}\phi_{\xi})=0 \label{237} \\
&&\phi_{\eta\xi}+\frac{1}{2}(G_{\eta}\phi_{\xi}+G_{\xi}\phi_{\eta})-
\frac{1}{T}(T_{\eta}\phi_{\xi}+T_{\xi}\phi_{\eta})=0 \label{238}
\end{eqnarray}
Also we can easily write Eqs  (\ref{223})  and  (\ref{224})  in the variables
introduced above.

One way of finding solutions is to solve the system of four Eqs
(\ref{236}) -  (\ref{238})  with four unknowns and then use Eqs
(\ref{223}) and  (\ref{224})  to get $ p $ and $\mu $. This form
of the system of equations i.e. four equations without $ p $ and
$\mu$ and two giving $ p $ and $\mu$ in term of the four dependent
variables is very convenient in the search for B\"{a}cklund
transformations \cite{ky:8}. Of course one can approach the
problem in a different way.
\section{Another Form of the System}
We shall write now the system we want to solve in a different way.
Eqs  (\ref{232})  and  (\ref{233})  become in the variables $\xi$ and
$\eta$ of Eqs  (\ref{234})
\begin{eqnarray}
&&G_{\eta\eta}+G^{2}_{\eta}-2G_{\eta}K_{\eta}+\frac{2}{(E+
\overline{E})^{2}}E_{\eta}{\overline{E}}_{\eta}=0 \label{31} \\
&&2K_{\eta\xi}-G_{\eta\xi}-G_{\eta}G_{\xi}+
\frac{1}{(E+\overline{E})^{2}}(E_{\eta}{\overline{E}}_{\xi}+
E_{\xi}{\overline{E}}_{\eta})=0 \label{32}
\end{eqnarray}
Let us put
\begin{equation}
G_{\eta}=N \label{33}
\end{equation}
which implies the relation $G_{\xi}=\overline{N}$ since G is real.
Then solving Eq  (\ref{31})  for $K_{\eta}$ and using the resulting
expression to eliminate $K_{\eta}$ from Eq  (\ref{32})  we get
\begin{eqnarray}
&&K_{\eta}=\frac{N_{\eta}}{2N}+\frac{N}{2}+
\frac{1}{(E+\overline{E})^{2}N}E_{\eta}{\overline E}_{\eta}
\label{34}\\
&&\frac{\partial}{\partial\xi}\{\frac{N_{\eta}}{N}+\frac{2}{(E+
\overline {E})^{2}N}E_{\eta}{\overline {E}}_{\eta}\}=N\overline
{N} \nonumber\\ &&-\frac{1}{(E+\overline
{E})^{2}}(E_{\xi}{\overline{E}}_{\eta}+
E_{\eta}{\overline{E}}_{\xi}) \label{35}
\end{eqnarray}
Also Eq  (\ref{220})  becomes
\begin{equation}
Im\Lambda=Im\{E_{\xi\eta}+\frac{1}{2}(NE_{\xi}+
{\overline {N}}E_{\eta})-\frac{2}{E+\overline {E}}E_{\xi}E_{\eta}\}=0
\label{36}
\end{equation}

The $ N $ we get by solving Eqs  (\ref{35}) and  (\ref{36})  must
be of the form of Eq  (\ref{33}).  But then we get
\begin{equation}
ImN_{\xi}=ImG_{\eta\xi}=0 \label{37}
\end{equation}
since $ G $ is real.We can show that any complex number $ N $
which satisfies the first of  Eqs  (\ref{37})  is of the form of
Eq (\ref{33})  with $ G $ real. Also Eq  (\ref{35})  implies the
relation
\begin{equation}
ImK_{\eta\xi}=0 \label{38}
\end{equation}
which means that $K_{\rho}$ and $K_{z}$ coming from Eqs  (\ref{34})
satisfy the integrability condition $K_{\rho z}=K_{z\rho}$. Using
Eq  (\ref{36})  we can write Eq  (\ref{35}) in the form
\begin{eqnarray}
&&(\frac{N_{\eta}}{N})_{\xi} - N{\overline N}+
\frac{2(E_n+\overline E_n)}{N(E+\overline E)^{2}} \{
E_{\xi\eta}+\frac{1}{2}(NE_{\xi}+{\overline
N}E_{\eta})\nonumber\\
&&-\frac{2}{E+\overline E}E_{\xi}E_{\eta}\}-
\frac{2}{(E+\overline E)^{2}}(\frac{N_{\xi}}{N^{2}}+
\frac{\overline N}{N})E_{\eta}{\overline E}_{\eta}=0 \label{39}
\end{eqnarray}
Therefore we can get $ N $ and $ E $ by solving
Eqs  (\ref{35})  or  (\ref{39}),  (\ref{36})  and the first of  (\ref{37})  and
 $ K $ from  (\ref{34}).

\end{document}